\documentstyle[12pt]{article}

\newcommand{\tr}{\mathop{\rm tr}\nolimits}
\normalbaselines
\catcode `\@=11
\@addtoreset{equation}{section}

\def\section{\@startsection {section}{1}{\z@}{-3.5ex plus -1ex minus
     -.2ex}{2.3ex plus .2ex}{\normalsize\bf}}
\def\subsection{\@startsection{subsection}{2}{\z@}{-3.25ex plus -1ex minus
-.2ex}{1.5ex plus .2ex}{\normalsize\bf}}

\def\thebibliography#1{\section*{References}
\list
  {[\arabic{enumi}]}{\settowidth\labelwidth{[#1]}\leftmargin\labelwidth
  \advance\leftmargin\labelsep
  \usecounter{enumi}}
  \def\newblock{\hskip .11em plus .33em minus -.07em}
  \sloppy
  \sfcode`\.=1000\relax}

\catcode `\@=12

\begin{document}

\begin{titlepage}
\vspace*{2.5cm}

\begin{center}

{\bf NEW METHODS IN THE THEORY OF GAUGE FIELDS}\vspace{1.3cm}\\
\medskip

{\bf Valery Koryukin}
\vspace{0.3cm}\\

 Department of Applied Mathematics, Mari State Technical University\\
 Lenin sq. --- 3, Yoshkar--Ola, 424024, Russia\\
 e-mail: koryukin@mpicnit.mari.su

\end{center}

\vspace*{0.5cm}

\noindent
 One of the fundamental problems of the theoretical physics is
 the search of the axioms, which ought to be the basis for the
 one-valued construction of Lagrangians of the relativistic
 fields. The creation of the gauge fields theory was the great
 success in the solution of this problem. The gauge formalism
 allowed to derive the total Lagrangians of the interacting fields
 from the postulated Lagrangians of the noninteracting (free)
 fields. We offer to do quite the reverse in consequence of what
 it is necessary to seek from the out set the construction principles
 of the total Lagrangians. By the theory construction we shall
 differ the wave-functions being the solutions of the
 differential equations (``theoretical'' functions) from the
 wave-functions which is constructed on the base of the experimental
 data possibly received by a scattering of particles (``empiric''
 functions). The ``empiric'' functions are necessary only for the
 definition (it is possibly only approximately) of the transition
 operators which will affect at the ``theoretical'' functions.
 This operators will be approximated the differential operators
 so, that the generalized variance of the differentiable
 ``theoretical'' fields will be the minimal one.



\end{titlepage}

\section{\hspace{-4mm}.\hspace{2mm} Introduction}

 As is known the experimentators spend many efforts to relieve
 of the useless information (of the noise). It can assume in this
 aspect what the human brain is achieving the more, in consequence of
 what a major proportion of the Universe is showed for us as an
 empty space with the Euclide properties, if any. We observe the
 presence of the matter only here and there so what it is necessary
 to ask:
 is the observable (of course also and by instruments) matter the
 main form or it play the off-beat role of the Brownian particles?
 This question became particularly a high-priority task since it
 was established the discrepancy of the galaxies kinetic energies
 and the galaxies potential energies with the virial
 theorem~\cite{ll}. Among the different hypothesis explaining this
 discrepancy we derive only those in which it was suggested to
 consider the neutrinos (under which we shall imply antineutrinos,
 too) as the fundamental form of the matter
 (it is necessary to note in consequence of this the Wheeler's
 book~\cite{wh}). In
 particular Pontekorvo and Smorodinskij~\cite{PS} explained the
 charge asymmetry of the baryon matter that it is only a slight
 fluctuation at the giant neutrinos background of the Universe.
 Having the neutrinos Universe and taking account of the
 Fermi-Dirac statistics we can remind about the Sakharov
 hypothesis~\cite{sah} in which the vacuum elasticity is connected
 with the gravitational interactions. As a result there is a chance
 to replace the vacuum elasticity by the neutrinos matter elasticity.

 So it was necessary
 to show that the gravitation interaction is not a fundamental
 one, but the one is induced by others interactions as possible
 hypothetical ones. The more so, that the gravitational
 constant~$G_N \approx 6.7 \cdot 10^{-39}~GeV^{-2}$ (it is used
 the system of units~$\hbar =c=1$,
 where~$2 \pi \hbar$ is the Planck's constant and~$c$ is the light
 speed) is a suspicionsly small value and a dimensional one
 furthermore (as is known the latter prevent to the constraction of
 the renormalizable quantum theory).
 Before building the theory of the induced gravitation on the base
 of the hypothetical interactions and the hypothetical particles
 it was necessary to verify the possibility of the utilization of
 the known particles and the known interactions for this purpose.
 Naturally that the neutrinos are the most suitable particles for
 this taking into account their penetrating ability, which allow
 them to interact with all the substance of the macroscopic
 body --- not with the surface layer only. As is known~\cite{GT},
 already in 30th Gamow and Teller offered to use the neutrinos for
 the explanation of the gravitation, but their mechanism provided
 the direct exchange of the pairs consisting of a neutrino and an
 antineutrino.

 Bashkin's works appearing in 80th on a propagation of the
 spin waves in the polarized gases~\cite{ba} allowed to make the
 supposition~\cite{k}, that the analogous collective oscillations are
 possible under certain conditions as well as in the neutrinos
 medium. If we shall consider that the effective temperature
 of the Universe neutrinos is the fairly low one then it is
 fulfilled one of conditions ($ \lambda \gg r_w$, where~$\lambda$
 is the de Broglie's wave-length of a neutrino and~$r_w$ is the
 weak interaction radius of an one~\cite{ba}) of the propagation of the
 spin wave in the polarized gases. As a result the quantum effects
 become the determing ones in such medium and the interference of
 the neutrinos fields (being the consequence of the known identity
 of elementary particles) must induce the quantum beats, which will
 be interpreted as zero oscillations of a vacuum. In consequence of
 this the mathematical apparatus~\cite{gm} applied by the
 description of the Casimir's effect~\cite{ca} can be used.

 We shall be interesting in quantum beats arising by the
 interference of the falling polarized flow of the Universe neutrinos
 on the macroscopic body with the scattered one at this
 body. Let's suppose for this the neutrinos have the zero rest
 mass (the other version~\cite{bv} will not be considered), so that
 the direction of their spin is connected hardly with the direction
 of their 3-velocity. In consequence of this only those neutrinos
 can be considered as ones forming the polarized flow,
 which propagate along straight line connecting specifically
 two particles of different macroscopic bodies. It explain the
 anisotropy of the zero quantum oscillations, which is necessary
 to  obtain  the right dependence~($1/R$) of the energy of the
 two-particles interaction on the distance~$R$ between particles
 in the Casimir's effect.

 Let's consider two macroscopic bodies with masses~$m_1$
 and~$m_2$ and with the fairly long distance~$R$ one from
 another~\cite{k1}. We shall  regard,  that the bodies
 contain~$2m_1 r_w$ and~$2m_2 r_w$ particles
 correspondingly (where the normalizing factor~$r_w$ is
 connected with cross-section~$\sigma_{\nu} $ of the neutrino upon
 the particle), implying
 thereby the statistics averaging of the properties of the
 elementary particles constituting the bodies.  If the
 particles of the macroscopic bodies had interacted with all
 neutrinos incidenting on them then these particles might have been
 considered as the opaque boandaries, which induce Casimir's
 effect on the straight line. By this the energy of the
 interaction of the particles would have been equal to~\cite{gm}
$$
\varepsilon_{AB} = \frac{1}{2}
  \sum_{n=1}^{\infty} \frac{\pi n}{R_{AB}} - \frac{1}{2}
  \int\limits_0^{\infty} \frac{\pi x}{R_{AB}} dx =
$$
\begin{equation}
 \label{1}
=\frac{i}{2} \int\limits_0^{\infty}
\frac{\pi (it)/R_{AB} - \pi (-it)/R_{AB}}{\exp (2\pi t) - 1}
 dt = - \frac{\pi}{24 R_{AB}}
\end{equation}
 ($A$ is a number of a particle of the first macroscopic
 body and~$B$ is a number of a particle of the second body).
 On account of the weakness of the  interaction of neutrinos
 with particles we are confined to a first approximation,  so that
 the energy~$E$ of the interaction of two macroscopic bodies is equal to
\begin{equation}
 \label{2}
 E\approx\sum_{A=1}^{2m_1 r_w}\sum_{B=1}^{2m_2 r_w}
 \varepsilon_{AB}.
\end{equation}
 Neglecting the dimensions of the bodies in comparison with interval
 $R$ between them (~$R_{AB}\approx R$), we shall  have finally
\begin{equation}
 \label{3}
 E\approx -2m_1 r_w~2m_2 r_w~\pi /(24R)
  = -G_N m_1 m_2 / R
\end{equation}
 where
\begin{equation}
 \label{4}
 G_N = \pi r_w^2 /6 = \sigma_{\nu} /6 .
\end{equation}

 If now we shall be based on the results of the experements fixing
 the equality of the gravitation mass and the inert one then it can
 consider that the spectrum of the particle masses is defined by
 their interaction with the neutrinos background of Universe. This
 statement is confirmed what the rest mass of the photon is equal to
 zero in contradistinction to the masses of the vector
 bosons $W^{+}$, $W^{-}$, $Z^{0}$ whiches interact with the neutrinos
 immediately~\cite{k2}.

 But the main idea is it now for us what the normal matter (not
 neutrinos) acts as the Brownian particles by the help of which it
 can make the attempt to estimate the statistic characteristics
 of the Universe neutrinos background. So from the formula (1.4)
 it can receive the estimate of the averaged neutrinos energy:
 $\omega_{\nu} \sim 10^{-13}~GeV$ \cite{k1}, what is sufficiently
 close to  the temperature ($\approx 2.3\cdot 10^{-13}~GeV $) of
 the Universe relict photons.

\section{\hspace{-4mm}.\hspace{2mm} The Lie local loop}

 We shall rely on the approach
 suggested by Schr\"odinger~\cite{S} which introduced
 the set of the unorthogonal to each other wave-functions
 describing the unspreading wave packet for a quantum
 oscillator. Later Glauber~\cite{G} showed a scope for a
 description of coherent phenomena in the optics by the
 Schr\"odinger introduced states and it was he who
 called them as coherent. This approach received the further
 development in Perelomov's work's who proposed the
 definition of the generalized coherent states specifically
 as the states arising by the action of the representation
 operator of a some transformation group on any fixed vector
 in the space of this representation~\cite{P}.

 It is what allow to give the physical interpretation to the gauge
 transformations by our opinion as the transformations inducing
 the generalized coherent states, which are characterized by the
 continuous parameters~\cite{Sh}. Admitedly if the parameters space
 are not the compact one (we shall consider the space-time manifold
 always as its subspace) then by the rather large changes of parameters
 it is necessary take into account the speed finity of the information
 propagation in consequence of what the coherence of the
 states are able to lose (what lead to the absence of the quantum
 phenomena on the macroscopic level). It makes us change the
 Perelomov definition considering it taken place for the arbitrary
 group only in the neighbourhood of identity, what gives rise to
 generalize the given definition not only for the Lie local groups but
 and for the Lie local loops.

 Let us to consider the wave packet the equivalence relation for the
 functions~$\Psi (\omega )$ of which we shall give by the
 infinitesimal transformations
\begin{equation}
 \label{1}
 \Psi \longrightarrow \Psi + \delta \Psi =
 \Psi + \delta T(\Psi ) ,
\end{equation}
 where~$\delta T$ is the particular case of the transition operator
 (at the begining the symmetry type  is not being specified)
 and~$\omega $ is a set of parameters characterizing the generalized
 coherent state.

 As the experiments on the scattering of particles in which the laws
 of the conservation are being prescribed are the sole source of the
 information about the structure of the space-time manifold on the
 microscopic level taking into account the Noether's theorem we
 introduce the finite-dimensional manifold~$M_r$ of
 parameters~$\omega^a$ ($a,b,c,d,e=1,2,...,r$), connecting its
 dimensionality~$r$ with the numbers of the conserved dynamical
 invariants. Further we shall consider the space~$M_r$ as the
 manifold, the parameters~$\omega^a$ as the coordinates of the
 point~$\omega \in M_r$ and we shall give the fields~$\Psi (\omega )$
 in a certain domain~$\Omega_r$ of the given manifold.
 We choose the arbitrary point~$e$ in the domain~$\Omega_r$
 and we shall consider that this point~$e$ is the centre of the
 coordinate system. Let the domain~$\Omega_r$ contain
 the subdomain~$\Omega_n$ with the point~$e$ by this the
 domain~$\Omega_n$ belong to a certain differentiable manifold~$M_n$
 (although it is possible in is convenient to define the
 manifold~$M_n$ separately from the manifold~$M_r$). Let moreover
 the set of the smooth curves belonging to the manifold~$M_n$
 have the common point~$e$. Define also the set of the vector
 fields~${\xi (x)}$ being tangents to this curves and we shall
 consider that a point~$x \in \Omega_n$ and on the domain~$\Omega_n$
 the own coordinate system is defined. It is convenient to give the
 differentiable manifold~$M_n$ by the differentiable
 functions~$\omega^a = \omega^a (x)$ in this case.

 Let~$\delta \Omega_r$ is the sufficiently small neighbourhood of the
 point~$e$, thereby and the sufficiently small
 neighbourhood~$\delta \Omega_n$ of the point~$x$ is being given
 ($x\equiv e\in \delta \Omega_n \subset \delta \Omega_r$). The
 coordinates of the point~$x$ note as~$x^i$ ($i,j,k,l,p,q=1,2,...,n$).
 Further we shall consider the fields~$\Psi (x)$ as the cross section
 of the vector fiber bundle~$E_{n+N}$. Using the vector
 fields~$\delta \xi (x)$ the coordinates of the neighbouring
 point~$x'=x+\delta x \in \delta \Omega_n$ are written down as
\begin{equation}
 \label{2}
 x'^i=x^i+\delta x^i \cong x^i+\delta \omega^a \xi_a^i (x) .
\end{equation}
 Comparing the values of the fields~$\Psi '(x)$ and~$\Psi (x'$), where
\begin{equation}
 \label{3}
 \Psi '(x') = \Psi + \delta\Psi = \Psi + \delta T(\Psi ) \cong
 \Psi + \delta\omega^a T_a(\Psi ) ,
\end{equation}
\begin{equation}
 \label{4}
 \Psi (x') =\Psi (x+\delta x) \cong \Psi +
 \delta \omega^a \xi_a^i \partial_i \Psi
\end{equation}
 ($\partial_i $ are the partial derivatives), we see that they are
 differing by the observables
\begin{equation}
 \label{5}
 \delta_o\Psi (x) \cong \delta\omega^a X_a(\Psi ) =
 \delta\omega^a [T_a(\Psi ) - \xi_a^i \partial_i\Psi ] ,
\end{equation}
 which can interpret as the deviations the field~$\Psi (x)$, received
 with the help of the transformations~(2.3). Further we shall consider
 the domain~$\delta \Omega_r \subset M_r$ as the domain of the Lie
 local loop~$G_r$ (specifically which can have and the structure of
 the Lie local group if we provide it with the property of the
 associativity) with the unit~$e$ induced by the set~$\{ T\}$,
 by this we shall consider the expression of the form~(2.3) as the
 infinitesimal law of the transformations of the Lie local loop of the
 fields~$\Psi (x)$. Precisely the structure of the Lie local loop will
 characterize the degree of the coherence considered by us the quantum
 system. By this the maximal degree is being reached for the Lie
 simple group and the minimal degree is being reached for the Abelian
 one. In the last case we shall have the not coherent mixture of the
 wave-functions, it's unlikely which can describe the unspreading
 wave packet that is being confirmed by the absence of the fundamental
 scalar particles, if hipothetical particles are not being taken into
 account (in experiments only the mesons, composed from the quarks,
 are being observed and which are not being considered the fundamental
 one). Note that the ``soft'' structure of the Lie local loop by
 contrast to the Lie group allow to use it by the description of the
 symmetry both the phase transition (there is the time dependence)
 and the compact objects (there is the space dependence) especially.

 As it's unlikely it can be to ignore an interaction between particles
 we must be able to select those interactions which interest us.
 Precisely therefore it makes sense to select the set of the operators
 which will play the role of the connection in further.
 Of course we take into account the dependence of the reference systems
 on the physical properties of the instruments (including the primary
 standards) and moreover what the part of the transition are not the
 observable ones. Let
\begin{equation}
 \label{6}
 L_a(\Psi ) = T_a(\Psi )+\xi_a^i \Gamma_i \Psi .
\end{equation}
 In consequence of this the formula~(2.6) is rewritten so
\begin{equation}
 \label{7}
 \delta_o\Psi \cong \delta\omega^a X_a(\Psi ) =
 \delta\omega^a [L_a(\Psi ) - \xi_a^i \nabla_i\Psi ] ,
\end{equation}
 where~$\nabla_i$ are the covariant derivatives with respect the
 connection~$\Gamma_i (x)$. Note, if~$L_a(\Psi ) = L_a\Psi $, then the
 following relations~\cite{K3}
\begin{equation}
 \label{8}
 \xi_a^i~\nabla_i\xi_b^k - \xi_b^i~\nabla_i\xi_a^k
 - 2~S_{ij}^k~\xi_a^i~\xi_b^j = - C_{ab}^c~\xi_c^k,
\end{equation}
\begin{equation}
 \label{9}
 L_aL_b - L_bL_a - \xi_a^i~\nabla_iL_a + \xi_b^i~\nabla_iL_a
 + R_{ij}~\xi_a^i~\xi_b^j = C_{ab}^c~L_c
\end{equation}
 must take place, where ~$S_{ij}^k(x)$ are the components of the
 torsion of the space-time~$M_n$
\begin{equation}
 \label{10}
 S_{ij}^k = (\Gamma_{ij}^k - \Gamma_{ji}^k)/2
\end{equation}
 and~$R_{ij}(x)$ are the components
 of the curvature of the connection~$\Gamma_i(x)$
\begin{equation}
 \label{11}
 R_{ij} = \partial_i \Gamma_j - \partial_j \Gamma_i +
 \Gamma_i \Gamma_j - \Gamma_j \Gamma_i .
\end{equation}
 Here and further~$\Gamma_{ij}{}^k(x)$ are the components of the
 internal connection of the space-time~$M_n$.
 By this the components~$C_{ab}^c(x)$ of the structural tensor
 of the Lie local loop~$G_r$ must satisfy to the identities
\begin{equation}
 \label{12}
 C_{ab}^c + C_{ba}^c = 0,
\end{equation}
\begin{equation}
 \label{13}
 C_{[ab}^d~C_{c]d}^e - \xi_{[a}^i~\nabla_{|i|} C_{bc]}^e
 + R_{ij[a}{}^e~\xi_b^i~\xi_{c]}^j = 0,
\end{equation}
 where~$R_{ija}{}^e(x)$ are the components of the curvature of the
 connection~$\Gamma_i{}_a^b(x)$
\begin{equation}
 \label{14}
 R_{ijb}{}^a = \partial_i \Gamma_j{}_b^a - \partial_j \Gamma_i{}_b^a +
 \Gamma_i{}_c^a \Gamma_j{}_b^c - \Gamma_j{}_c^a \Gamma_i{}_b^c .
\end{equation}

 We construct the differentiable manifold~$M_n$, not interpreting it
 by physically. Of course we would like to consider the manifold~$M_n$
 as the space-time~$M_4$. At the same time it is impossible to take
 into account the possibility of the phase transition of a system as a
 result of which it can expect the appearance of the coherent states.
 In consequence of this it is convenient do not fix the dimensionality
 of the manifold~$M_n$. It can consider that the macroscopic system
 reach the precisely such state by the collapse. As a result we have
 the classical analog of the coherent state of the quantum system.
 Besides there is the enough developed apparatus --- the dimensional
 regularization using the spaces with the changing dimensionality and
 representing if only on the microscopic level.

\section{\hspace{-4mm}.\hspace{2mm} The gauge fields}

 We shall demand the minimality of the variations~(2.5) (or~(2.7)), if
 only on the ``average'', in order to can be hope for the set of the
 fields~$\Psi (x)$ is capable to describe the wave packet.
 Consider for this the following integral
\begin{equation}
\label{1}
 {\cal A} = \int\limits_{\Omega_n} {\cal L} d_nV =
 \int\limits_{\Omega_n} \kappa \overline{X^b} (\Psi)
 \rho_b^a X_a (\Psi) d_n V ,
\end{equation}
 being the analogue of the fields~$\Psi (x)$ variance in the
 domain~$\Omega_n$ at issue, which we shall call the action,
 and~${\cal L}$ we shall call the Lagrangian. Here and further
 $\rho_a^b (x)$ are the components of the density matrix~$\rho (x)$
 (note that Latin indexis are the only (possible) visible part of the
 indexis of the density matrix,  $\tr \rho = 1, \rho^+ = \rho$, the
 top index~$+$ is the symbol of the Hermitian conjugation), and the
 bar means the Dirac conjugation which is the
 superposition of the Hermitian conjugation and the space inversion.
 Solutions~$\Psi (x)$ (and even one solution) of equations, which are
 being produced by the requirement of the minimality of the
 integral~(3.1) can be used for the constraction of the all set of the
 functions~$\{ \Psi (x)\}$ (generated by the transition operator),
 describing the wave packet. It is naturally
 to demand the invariance of the integral~(3.1) relatively the
 transformations~(2.2) and~(2.3), in consequence of what it is
 necessary to introduce the additional fields~$B(x)$ with the
 transformation law in point~$x \in \delta \Omega_n $ in the form
\begin{equation}
\label{2}
 \delta_o B = \delta\omega^a Y_a(B) +
 \nabla_i\delta\omega^a Z_a^i(B) ,
\end{equation}
 and which we shall name the gauge ones. Make it in the standard
 manner defining them by the density matrix~$\rho (x)$ as
\begin{equation}
\label{3}
 B_{\gamma}^b \overline{B^{\gamma}_a} =
 \rho_a^b (B_{\gamma}^c \overline{B^{\gamma}_c}) .
\end{equation}
 by this the factorization of the gauge fields~$B(x)$ on
 equivalence classes is allowed for the writing of the indexes
 of their components~$B_{\alpha}^a(x)$.
 Note, that~$B_{\alpha}^a(x)$ can be both Utiyama gauge
 fields~\cite{ut} and Kibble gauge fields~\cite{kib}.
 Following for Utiyama~\cite{ut} we shall not concretize
 significances which are adopted by the Greek indexes.

 Further we shall assume that the density matrix~$\rho (x)$ defines
 the dimensionality of manifold~$M_n$, using even if for this the
 corresponding generalized (singular) functions in consequence of
 what the rank
 of the density matrix~$\rho (x)$ must be equal to~$n$, and the
 formula~(3.1) can be rewritten in the form
\begin{equation}
\label{4}
 {\cal A} = \int\limits_{\Omega_r} {\cal L} d_rV =
 \int\limits_{\Omega_r} \kappa \overline{X^b} (\Psi)
 \rho_b^a X_a (\Psi ) d_r V ,
\end{equation}
 We should connect the rank~$n$ of the density matrix~$\rho (x)$ with
 the nonzero vacuum average of the gauge fields~$B_{\alpha}^a$.

 In consequence of (3.1) and (3.3) it can consider that the
 Lagrangian~${\cal L}$ depend on the gauge fields~$B$ by
\begin{equation}
\label{5}
 D_{\beta}\Psi =-B_{\beta}^aX_a(\Psi),
\end{equation}
 We see by this formula, that the particles charges define the form
 of the generators~$X_a(\Psi)$ and the structural tensor
 components~$C_{ab}^c(x)$ of the Lie local loop~$G_r$, and hence it
 follows the dependence of the symmetries on the particles charges
 as and in the Utiyama formalism~\cite{ut}. If now we shall have
 ``spread'' the gauge fields but retaining the terms~$L_i$
 responsible for the vacuum oscillation, then the Lagrangian~${\cal L}$
 will have been rewritten in the form
\begin{equation}
\label{6}
 {\cal L} = k (\partial_i\overline{\Psi} - \overline{L_i\Psi} )
 \eta^{ij} (\partial_j\Psi - L_j\Psi ) .
\end{equation}

 In particular by~$n=4$ and considering the CPT--degeneracy
\begin{equation}
\label{7}
 \Psi = \left(\begin{array}{c}
 \psi_L\\
 \psi_R
 \end{array}  \right) , \quad
 \psi_L = \frac12 (I - \gamma_5) \psi, \quad
 \psi_R = \frac12 (I + \gamma_5) \psi, \quad
 \psi = \left(\begin{array}{c}
 \psi_1\\
 \psi_2\\
 \psi_3\\
 \psi_4
 \end{array}  \right) ,
\end{equation}
\begin{equation}
\label{8}
 L_A = - L^+_A = i \frac{\omega_o}{2}
 \Sigma_A \otimes \left(\begin{array}{cc}
                        I - \gamma_4 & 0 \\
                        0 & I + \gamma_4
                        \end{array}\right), \quad
                       \Sigma_A = \left(\begin{array}{cc}
                                          \sigma_A & 0 \\
                                            0 & \sigma_A
                      \end{array}\right),
\end{equation}
\begin{equation}
\label{9}
 L_4 = - L^+_4 = i \frac{\omega_o}{2}
 \gamma_4 \otimes \left(\begin{array}{cc}
                 I - 3\gamma_4 & 0 \\
                 0 & I + 3\gamma_4
                 \end{array}\right)
\end{equation}
 ($i^2 = - 1$; $I$~is the unit matrix;
 $\gamma_5 = - i \gamma_1 \gamma_2 \gamma_3 \gamma_4$;
 $\gamma_1, \gamma_2, \gamma_3, \gamma_4$ are the Dirac
 matrices; $\sigma_A$ are the Pauli matrices; $A,B = 1,2,3$;
 $\eta^{ij}$ are the contravariant components of the
 metric tensor of the Minkowski space; $\omega_o $ is a constant)
 it can obtain the
 Lagrangian~${\cal L}$ of the neutrinos fields in the standard
 form ($\omega_o = 1/k$)
\begin{equation}
\label{10}
 {\cal L} = -i k
 \frac{\omega_o}{2} [\eta^{AB} (\partial_A\overline{\psi}\gamma_B\psi
 - \overline{\psi} \gamma_A \partial_B\psi ) -
 \partial_4\overline{\psi} \gamma_4 \psi +
 \overline{\psi} \gamma_4 \partial_4\psi )]
\end{equation}
 If now we turn on the mixing of the fields with different
 polarization, it is possible substituting~$L_4$ (the formula
 (3.9)) in the following form
\begin{equation}
\label{11}
 L_4 = - L^+_4 = i \frac{\omega_o}{2}
 \gamma_4 \otimes \left(\begin{array}{cc}
                 (I - 3\gamma_4 ) & 2I\omega_1 /\omega_o \\
                 2I\omega_1 /\omega_o & (I + 3\gamma_4 )
                 \end{array}\right),
\end{equation}
 then the fields~$\Psi (x)$ will describe the particles with
 the nonzero rest masses. Of course this mixing is the detector of
 the vacuum frequency change, which is induced by the presence
 of the added fields (in particular, by the presence of the
 electromagnetic field).

 Now one may proceed to a construction of the covariant gauge
 formalism considering that the manifold~$M_n$ is the
 Riemannian space-time. For  this it is necessary to find a law
 of a transformation of the fields~$B(x)$. Let the
 fields~$D_{\alpha} \Psi$ change analogously to the
 fields~$\Psi(x)$ in a point~$x\in M_n$, then is
\begin{equation}
\label{12}
 \delta_0 D_{\alpha}\Psi =\delta\omega^b~(L_bD_{\alpha}\Psi
 - L_b{}_{\alpha}^{\beta}~D_{\beta}\Psi -
 \xi_b^i~\nabla_i D_{\alpha}\Psi).
\end{equation}
 As a result~$\delta_0 B_{\alpha}^a$ are written down in the form:
\begin{equation}
\label{13}
 \delta_0 B_{\alpha}^d =
 \delta\omega^b~(C_{cb}^d~B_{\alpha}^c -
 L_b{}_{\alpha}^{\beta}~B_{\beta}^d -
 \xi_b^i~\nabla_i B_{\alpha}^d) +
\Phi_{\alpha}^i~\nabla_i \delta\omega^d ,
\end{equation}
where
\begin{equation}
\label{14}
 \Phi_{\beta}^i = B_{\beta}^a~\xi_a^i,
\end{equation}

 Since the action
\begin{equation}
\label{15}
 {\cal A}_t =
 \int\limits_{\Omega_n} {\cal L}_t~\eta~dx^1 dx^2 ... dx^n
\end{equation}
 ($\Omega_n$ is a region of the space-time~$M_n$
 and~$\eta (x)$ is the base density of the same) must be invariant
 against infinitesimal transformations of the Lie local
 loop~$G_r$, then the total Lagrangian~${\cal L}_t$ depending on
 fields~$\Psi (x)$,~$B(x)$ and also their derivatives of
 the first order is unable to be selected arbitrarily. The
 following Lagrangian~${\cal L}_t(\Psi ;
 D_{\alpha}\Psi; F_{\alpha\beta}^c)$ satisfy to this demand,
 where the components~$F_{\alpha\beta}^c(x)$ of the intensities
 of the gauge fields~$B(x)$ have the form:
$$
 F_{\alpha\beta}^c =[\delta_b^c -
 \xi_b^i~\Phi_i^{\gamma}~(B_{\gamma}^c -
 \beta_{\gamma}^c)]~[\Phi_{\alpha}^j~\nabla_j B_{\beta}^b -
 \Phi_{\beta}^j~\nabla_j B_{\alpha}^b -
$$
\begin{equation}
\label{16}
 B_{\alpha}^e~B_{\beta}^d~C_{ed}^b +
 (B_{\alpha}^e~L_{e\beta}{}^{\delta} -
 B_{\beta}^e~L_{e\alpha}{}^{\delta})~B_{\delta}^b].
\end{equation}
 Note that the fields~$\Phi_i^{\alpha} (x)$ are defined
 from the equations: $\Phi_{\alpha}^i~\Phi_j^{\alpha} =\delta_j^i$
 ($\delta_i^j$ and~$\delta_a^b$ are the Kronecker delta symbols).
 The components~$\beta_{\alpha}^b (x)$ of linear homogeneous
 geometrical objects are being interpreted as vacuum averages of
 gauge fields~$B(x)$. In consequence of~$\beta_{\alpha}^b (x) \ne 0$
 the matrixes~$L_i$ in the formula~(3.6) proved to be the non-zero ones.

 Rewrite the equations
\begin{equation}
\label{17}
 \Phi_{\alpha}^i~\left(\frac{{\cal L}_t}{\eta}~\frac{\partial
 \eta}{\partial B_{\alpha}^b} + \frac{\partial {\cal L}_t}
 {\partial B_{\alpha}^b} -
 \nabla_j\left(\frac{\partial {\cal L}_t}
 {\partial\nabla_j B_{\alpha}^b}\right)\right) = 0
\end{equation}
 of gauge fields in the quasi-maxwell form:
\begin{equation}
\label{18}
 \nabla_j H_a^{ji} = I_a^i,
\end{equation}
 where
\begin{equation}
\label{19}
 H_a^{ij} = -\Phi_{\beta}^i~\frac{\partial{\cal L}_t}
 {\partial\nabla_j B_{\beta}^a} = \Phi_{\beta}^j~
 \frac{\partial{\cal L}_t}{\partial\nabla_i B_{\beta}^a},
\end{equation}
\begin{equation}
\label{20}
 I_a^i = - {\cal L}_t\xi_a^i -
 \frac{\partial {\cal L}_t}{\partial\nabla_i\Psi}~X_a(\Psi) -
 \frac{\partial {\cal L}_t}{\partial\nabla_i B_{\beta}^b}~
 Y_{a\beta}^b(B),
\end{equation}
\begin{equation}
\label{21}
 Y_{a\gamma}^b(B) = C_{ca}^b~B_{\gamma}^c -
 L_b{}_{\alpha}^{\beta}~B_{\beta}^d -
 \xi_a^i~\nabla_i B_{\gamma}^b.
\end{equation}
 Besides let
\begin{equation}
\label{22}
 \Phi_{\alpha}^{k} \frac{\partial\eta}{\partial B_{\alpha}^b} +
 \eta \xi_b^k = 0.
\end{equation}
 We pick out from the equations of gauge fields folding them
 with~$B_{\alpha}^b~\Phi_l^{\alpha}$ those which can will be
 called the equations of fields~$\Phi_{\alpha}^i (x)$ and which
 must substitute for Einstein gravitational equations.

 Furter we shall consider the making of most important equations
 within the scope of the offered gauge field theory. Note that the
 base physical workload will impinge on the density matrix, the
 choice of which will be defined by the symmetries of the physical
 system states.

\section{\hspace{-4mm}.\hspace{2mm} The Einstein gravitational equations}

 Let $n=4$ and the Greek indexes take the values~$1,2,3,4$. Let
 moreover $\eta_{\alpha\beta}$ are the covariant components
 and $\eta^{\alpha\beta}$ are the contravariant components of the
 metric tensor of the Minkowski space. Introduce the geometrical
 objects by definitions
\begin{equation}
\label{1}
 h_{\alpha}^i = \beta_{\alpha}^a \xi_a^i,
 \quad \eta^{ij}=\eta^{\alpha\beta} h_{\alpha}^i h_{\beta}^j,
 \quad g^{ij}=\eta^{\alpha\beta}\Phi_{\alpha}^i\Phi_{\beta}^j
\end{equation}
 and define also the geometrical objects $h_{\alpha}^i(x)$,
 $\eta_{ij}$, $g_{ij}$ as the solutions of the equations
\begin{equation}
\label{2}
 h^{\alpha}_k h_{\alpha}^i = \delta_k^i,
 \quad \eta^{ij} \eta_{kj} = \delta_k^i,
 \quad g^{ij} g_{kj} = \delta_k^i.
\end{equation}
 We shall consider that $g_{ij}$ are the covariant components of the
 metrical tensor of the Riemannian space-time~$M_4$ in consequence
 of what
\begin{equation}
\label{3}
 \nabla_k g_{ij}=0,\quad \nabla_k g^{ij}=0.
\end{equation}
 As a result the construction of the differentiable manifold~$M_4$
 can be connected with the finding of the equations solutions of
 the gauge fields~$\Phi_{\beta}^i = B_{\beta}^a \xi_a^i $ received
 from the demand of the minimality of the total action
\begin{equation}
\label{4}
 {\cal A}_t = \int\limits_{\Omega_4} {\cal L}_t d_4V =
 \int\limits_{\Omega_4} [{\cal L}(\Psi, D_{\alpha}\Psi) +
 {\cal L}_1(F_{\alpha\beta}^a)] d_nV ,
\end{equation}
 where (in particular $k_1 = 1/(16\pi G_N)$)
\begin{equation}
\label{5}
 {\cal L}_1(F_{\alpha\beta}^a) = \kappa_1
 \overline{F^{\beta \delta}_d}
 \rho_1{}_{\beta \delta}^d{}^{\alpha \gamma}_b F_{\alpha \gamma}^b.
\end{equation}

 It is naturally that the supposition about fields and particles (the
 neutrinos) filling the Universe and defining the geometrical structure
 of the space-time manifold, allow to introduce the connection of the
 fundamental tensor of this manifold with that kind of the statistical
 characteristic as the entropy defining it in a standard manner by the
 reduced density matrix~$\rho' (x)$ in the form
\begin{equation}
\label{6}
  S = - \tr (\rho' \ln \rho' ) ,
\end{equation}
 where the components of the reduced density matrix~$\rho '(x)$ are
 defined as
\begin{equation}
\label{7}
 \rho_i^j = \overline{\xi_i^b}\rho_b^a\xi_a^j /
 (\overline{\xi_k^d}\rho_d^c\xi_c^k) .
\end{equation}
 As a result the transition from the singular state of the Universe to
 the modern one with the non-zero vacuum averages~$\beta^b_{\alpha}$
 of the gauge fields~$B^b_{\alpha}$ must be defined by the growth
 of the entropy~$S$.

 Further we shall consider for a simplification of a calculation that
 the Lie transitiv local loop~$G_r$ act effectively in the considered
 domain of the space-time~$M_4$, in consequence of this~$r=4$ and let
\begin{equation}
\label{8}
 L_c{}_{\alpha}^{\beta} = 0 .
\end{equation}
 As a result the formula~(3.21)
 is rewritten as
\begin{equation}
\label{9}
 Y_{a\gamma}^b(B)\longmapsto Y_{i\gamma}^k = -\nabla_i\Phi_{\gamma}^k .
\end{equation}
 Write down the Lagrangian~${\cal L}_1$ in the form
\begin{equation}
\label{10}
 {\cal L}_1=\kappa_1\eta^{\alpha\beta}
 Y_{j\beta}^i(\Phi) Y_{i\alpha}^j(\Phi) ,
\end{equation}
 considering that the fields~$\Phi (x)$ satisfy the Lorentz
 conditions:
\begin{equation}
\label{11}
 \nabla_i\Phi_{\alpha}^i = 0
\end{equation}
 and~$\eta^{\alpha\beta}$ are the contravariant components of the
 metric tensor of the Minkowski space. In this case the
 Lagrangian~${\cal L}_1$ is distinguished only the constant
 factor~$(-\kappa_1)$ from the scalar curvature~$R=g^{ij}R_{ij}$
 where~$R_{ij}=R_{kij}{}^k$ are the components of the Ricci tensor.
 It can note that in this case the gauge fields equations are
 written down as the Einstein equations, namely
\begin{equation}
\label{12}
 I_k^j=\kappa_1(2g^{jl}R_{kl}-\delta_k^jR).
\end{equation}
 where the energy-momentum tensor of the everybody fields
 (excluding the fields~$\Phi_{\alpha}^i$) has the form
\begin{equation}
\label{13}
 I_k^j = - \delta_k^j {\cal L} +
 \frac{\partial {\cal L}}{\partial \Phi_{\alpha}^k}
 \Phi_{\alpha}^j .
\end{equation}

 If we do not wish to use the Lorentz' conditions (4.11), then it
 can take the following Lagrangian~${\cal L}_1$
\begin{equation}
\label{14}
 {\cal L}_1=\frac12\kappa_1\eta^{\alpha\beta}
 \eta_{ijpq}\eta^{klpq}
 Y_{k\beta}^i(\Phi) Y_{l\alpha}^j(\Phi) ,
\end{equation}
 where~$\eta^{klpq}$ is the base $4$-vector of the manifold~$M_4$
 and~$\eta_{ijpq}$ is the mutual one to~$\eta^{klpq}$. As a result
\begin{equation}
\label{15}
 {\cal L}_1=\kappa_1\eta^{\alpha\beta}
 (\nabla_j \Phi_{\beta}^i\nabla_i \Phi_{\alpha}^j -
 \nabla_i \Phi_{\beta}^i\nabla_j \Phi_{\alpha}^j),
\end{equation}
 It can rewrite this Lagrangian (4.15) also in the form
\begin{equation}
\label{16}
 {\cal L}_1=\frac14\kappa_1\eta^{\alpha\beta}
 (F_{\gamma\beta}^{\delta} F_{\mu\alpha}^{\nu}
 \eta^{\gamma\mu}\eta_{\delta\nu} +
 2F_{\gamma\beta}^{\delta} F_{\delta\alpha}^{\gamma} -
 4F_{\gamma\beta}^{\gamma} F_{\delta\alpha}^{\delta} )
\end{equation}
 where the components~$F_{\mu\alpha}^{\nu}$ of the intensities
 of the gauge fields~$\Phi (x)$ can be got from the
 intensities~(3.16) as
\begin{equation}
\label{17}
 F_{\alpha\beta}^{\nu} = \Phi_k^{\nu}
 (\Phi_{\alpha}^i\nabla_i\Phi_{\beta}^k -
 \Phi_{\beta}^i\nabla_i\Phi_{\alpha}^k ) .
\end{equation}

 In the more general case, when~$r\ge 4$, it ought to use the
 following total Lagrangian~${\cal L}_t$
 (in particular $k_1 = 1/(16\pi G_N), k_2 = 1/(4\pi) $)
$$
 {\cal L}_t = F_{\alpha\beta}^a F_{\gamma\delta}^b
 \eta^{\beta\delta}[\kappa_1\xi_a^i\xi_b^j
 (\eta^{\alpha\gamma} \eta_{\varepsilon\theta}
 h_i^{\varepsilon} h_j^{\theta} + 2 h_i^{\gamma} h_j^{\alpha}
 - 4 h_i^{\alpha} h_j^{\gamma}) +
$$
\begin{equation}
\label{18}
 \kappa_2 \eta_{cd} \eta^{\alpha\gamma}
 (\delta_a^c - \xi_a^i h_i^{\varepsilon}\beta_{\varepsilon}^c)
 (\delta_b^d - \xi_b^j h_j^{\theta} \beta_{\theta}^d) ] /4
 + {\cal L}(\Psi, D_{\alpha}\Psi) ,
\end{equation}
 where $\eta_{ab}(x)$ are the components of the symmetric
 undegenerate tensor fields. Now the Einstein equations, got from
\begin{equation}
\label{19}
 \Phi_{\alpha}^j~\left(\frac{{\cal L}_t}{\eta}~\frac{\partial
 \eta}{\partial B_{\alpha}^b} + \frac{\partial {\cal L}_t}
 {\partial B_{\alpha}^b} -
 \nabla_i\left(\frac{\partial {\cal L}_t}
 {\partial\nabla_i B_{\alpha}^b}\right)\right)
 B_{\beta}^b \Phi_k^{\beta} = 0 ,
\end{equation}
 are written as
\begin{equation}
\label{20}
 g^{ji} R_{ki} - \frac12 \delta_k^j R =
 \frac{1}{2\kappa_1} [D_a^{ij} E_{ik}^a -
 \frac14 \delta_k^j D_a^{il} E_{il}^a +
 P^j \Psi D_k \Psi - \delta_k^j {\cal L}(\Psi, D_i\Psi)] ,
\end{equation}
 where
\begin{equation}
\label{21}
 D_i \Psi = \Phi_i^{\alpha} D_{\alpha}\Psi =
 \nabla_i\Psi - B_i^a L_a\Psi ,
\end{equation}
\begin{equation}
\label{22}
 P^k\Psi = \frac{\partial {\cal L}}{\partial D_k\Psi} =
 \Phi_{\alpha}^k\frac{\partial {\cal L}}{\partial D_{\alpha}\Psi} ,
\end{equation}
\begin{equation}
\label{23}
 B_i^a = \Phi_i^{\alpha} B_{\alpha}^a ,
\end{equation}
\begin{equation}
\label{24}
 E_{ij}^a = (\delta_b^a - \xi_b^k B_k^a) (\nabla_i B_j^b -
 \nabla_j B_i^b + B_i^c B_j^d C_{cd}^b) ,
\end{equation}
\begin{equation}
\label{25}
 D_a^{ij} = \kappa_2 g^{ik} g^{jl} \eta_{cd}
 (\delta_a^c - \xi_a^p h_p^{\varepsilon}\beta_{\varepsilon}^c)
 (\delta_b^d - \xi_b^q h_q^{\theta} \beta_{\theta}^d) E_{kl}^b .
\end{equation}
 By this it can rewrite the total Lagrangian~${\cal L}_t$ (4.18) as
\begin{equation}
\label{26}
 {\cal L}_t = (H_k^{ij} F_{ij}^k + D_a^{ij} E_{ij}^a)/4
 + {\cal L}(\Psi, D_{\alpha}\Psi) ,
\end{equation}
 where
\begin{equation}
\label{27}
 F_{ij}^k = - \Phi_{\alpha}^k(\nabla_i\Phi_j^{\alpha}
 - \nabla_j\Phi_i^{\alpha} )
\end{equation}
and
\begin{equation}
\label{28}
 H_k^{ij} = \kappa_1 (g_{kl} F_{pq}^l g^{ip} g^{jq} +
 F_{kl}^i g^{jl} + g^{il} F_{lk}^j + 2 g^{ip} \delta_k^j F_{lp}^l
 + 2 \delta_k^i g^{jp} F_{pl}^l) .
\end{equation}

 In the general case the condition~(4.8) it is necessary to abolish
 ($L_c{}_{\alpha}^{\beta}\ne 0$) so that the
 intensities~$F_{\alpha\beta}^{\nu}$ of the fields~$\Phi (x)$ will have
 the following form
\begin{equation}
\label{29}
 F_{\alpha\beta}^{\nu} = \Phi_k^{\nu} (\Phi^i_{\alpha}
 \nabla_i\Phi^k_{\beta} - \Phi^i_{\beta} \nabla_i\Phi^k_{\alpha}) +
 B^c_{\alpha} L_c{}_{\beta}^{\nu} - B^c_{\beta} L_c{}_{\alpha}^{\nu} .
\end{equation}
 Already because of this the masses of the vector
 bosons (being the quanta of the gauge fields)
 can be the non-zero ones. Thus the interactions of the
 elemental particles with fields~$\Phi_{\alpha}^i$ which are
 describing the vacuum oscilations and which are connected with the
 gravitational interactions can lead to the appearance of the masses
 both of the fermions and of the vector bosons.

\end{document}